# DLT Compliance Reporting

*Research in progress*

Henrik Axelsen
University of Copenhagen
heax@di.ku.dk

Johannes Rude Jensen
University of Copenhagen
eToroX Labs
johannesrudejensen@gmail.com

Omri Ross
University of Copenhagen
eToroX Labs
Omri@di.ku.dk

## Abstract

The IS discourse on the potential of distributed ledger technology (DLT) in the financial services has grown at a tremendous pace in recent years. Yet, little has been said about the related implications for the costly and highly regulated process of compliance reporting. Working with a group of representatives from industry and regulatory authorities, we employ the design science research methodology (DSR) in the design, development, and evaluation of an artefact, enabling the automated collection and enrichment of transactional data. Our findings indicate that DLT may facilitate the automation of key compliance processes through the implementation of a 'pull-model', in which regulators can access compliance data in near real-time to stage aggregate exposures at the supranational level. Generalizing our preliminary results, we present four propositions on the implications of DLT in compliance. The findings contribute new practical insights on the topic of compliance to the growing IS discourse on DLT.

## Introduction

All public and private companies operating in developed economies are subject to some level of regulatory compliance, either in the business reporting context, or through requirements for financial accounting. Due to the systemic importance of large financial institutions in the global economy, banks are amongst the most heavily regulated organizations and subject to strict compliance reporting requirements, ranging from data gathered for the compilation of macroeconomic statistics all the way down to microeconomic supervisory needs. Since the financial crisis in 2008, regulatory reporting requirements within the EU has grown by more than 40 pieces of legislation. This has generated a significant number of new and granular reporting requirements imposing additional pressure on both authorities' and financial institutions' reporting systems (Baudino et al. 2020).

Reporting obligations are specified at the global level through supranational bodies such as the Bank for International Settlement (BIS) and the Basel Committee on Banking Supervision (BCBS) and transposed to the European level in a variety of legal frameworks. Frameworks span from the macro-level mandated the European System of Central Banks' Integrated Reporting Framework, down to the micro-level supervision tasks mandated by EU directives and regulations that are interpreted by the European Banking Authority (EBA). These international bodies mandate prudential risk reporting through *Implementing Technical Standards* (ITS), the compliance reporting obligation we focus on in this short paper, that require local supervisors to collect aggregated risk data from banks. Not counting significant ad hoc requirements, the ITS risk reporting comprises more than 500 complex reporting obligations, incorporating thousands of tables and containing tens of thousands of data fields, the combination of which is used to produce different kinds of reports submitted to regulators on a monthly, quarterly, semi-annual or annual basis. The annual cost of ITS risk reporting is estimated at up to €12bn annually for the population of around 5,000 banks in the EEA, equivalent to approximately one third of banks' total cost of compliance (Eba 2021; EBA 2021a; European Commission 2021b). In practice, it is the banks that collect data from their internal systems, map this operational data to the data elements needed to populate regulatory reports (so-called 'input data'), transforming reporting data based on reporting instructions and subsequently submitting reports to the competent authorities. Because banks are responsible for submitting this data themselves, this model is known as the *push model*. European banks have made moderate progress in improving data management in the *push model* since facing strict obligations to do so since 2013. Yet, material challenges remain unsolved across markets, mainly due to a lack of alignment between new IT solutions and legacy systems (BCBS 2020). These technical challenges are exacerbated by an increasingly complex regulatory environment in which regulators frequently introduce changes to reporting frameworks and require multiple different data models for different ITS reporting requirements. As a result, banks often take up to 90 days to produce compliance reports, even under stressed conditions. This can have highly detrimental implications to regulators ability to understand systemic and structural risks to the European economies.

Working with a group of eight stakeholders representing perspectives from banking, central banking, supervisory authorities, and banking regulators within the European context, we examine how the DLT-based solutions emerging between institutions and governmental bodies could reduce the reporting burden, by facilitating a so-called *pull-model* for compliance data, enabling regulatory bodies to *pull* any the necessary data as it is produced in real time. We address the research question: *To what extent could the adoption of DLT based solutions optimize ITS compliance reporting for banks and organizations in the EEA?* We present ongoing work towards the design of a DLT agnostic artefact designed to collect and enrich transaction data with ITS reporting compliance data. While the IS discourse on the efficacy and potential of DLT in financial processes has grown at a tremendous pace in recent years, little has been said about the implications that the adoption of DLT will introduce for compliance reporting. By employing the DSR methodology in the design and evaluation of a conceptual artefact with a large group of stakeholders at international and governmental institutions, we seek to contribute new practical and actionable insights on the topic of compliance to the growing DLT discourse in IS.

## Compliance Reporting and Distributed Ledger Technology (DLT) in IS

In the 'push-model' for compliance reporting (Figure 1.), local banks push data to local authorities, which subsequently consolidate banking group reports and push these to the supranational level. The nationally competent authorities (NCA) for supervision, resolution (NRA) and central banking (NCB) are subsequently responsible for pushing the data forward to the respective targets at the European Banking Authority (EBA), the European Central Bank (ECB) and the Single Resolution Board (SRB) at the European level. The resolution authorities are included in the flow as they cooperate with the other institutions and the reporting obligations are used to assess whether a bank is failing or likely to fail and how to resolve it.

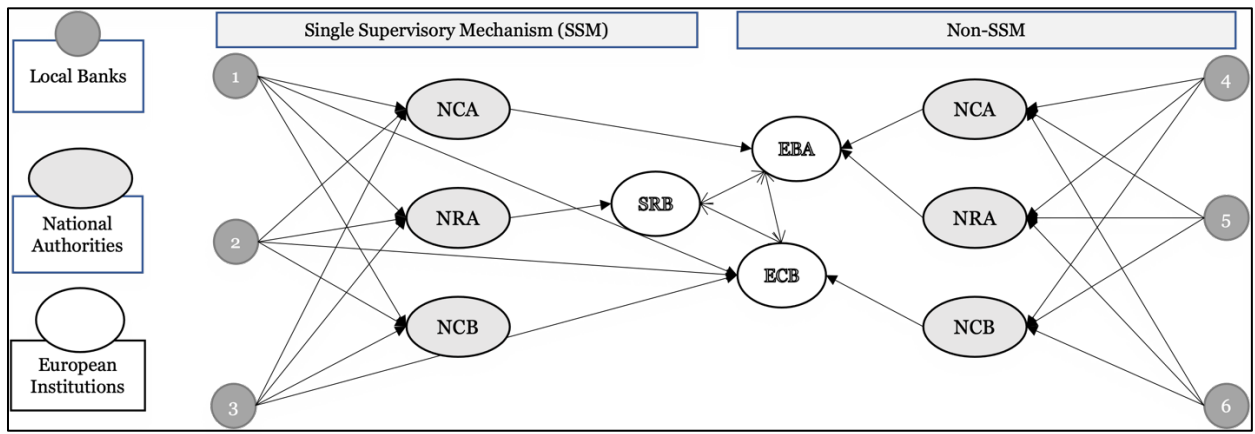

**Figure 1: Illustration of the Current 3-tier push-model for ITS compliance reporting**

Practically, reporting is initiated by the local banks, that submit a pre-defined XML-report or create it manually and submit to the local authorities, often through a jointly operated portal by the NCB and NCA. As the reporting obligations include sensitive data related to privacy, banking regulatory secrecy and competitive status, data is masked for analytical purposes and further truncated, so that sensitive data is not easily identifiable. The data is subsequently processed to create supervisory or statistical reporting, which is pushed to the European authority level as stipulated by the supranational bodies policy mandates. Traditionally, data security is managed via identity and access management controls, network segmentation, strong communication protocols supported by firewalls, data segregation, monitoring, and process controls to avoid leakage and abuse. Figure 2 shows the steps of the compliance reporting process (EBA 2021b).

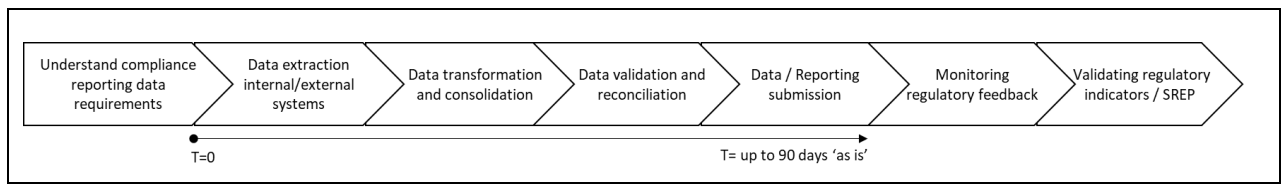

**Figure 2: Compliance reporting process**

Distributed Ledger Technology (DLT) denotes a distributed transactional database that is replicated across multiple peers in a network with a shared communication protocol, facilitating a tamper-proof record of transactions (Beck and Müller-bloch 2018; Glaser 2017). In recent years, IS scholars have demonstrated how DLT may (I) enable atomic settlement of transactions (Ross et al. 2019; Wüst and Gervais 2017) and automate the execution of OTC derivatives (Jensen and Ross 2020); (II) increase resiliency (no 'single point of failure') while reducing ambiguity in transactions by providing full disclosure of a 'single truth' for all network participants (Nærland et al. 2018); (III) Simplify and automate collection, sharing, reconciliation and reporting processes for sensitive data (Moyano and Ross 2017) while increasing transparency and reducing operational risk (Cerchiaro et al. 2021); (IV) fully interoperable and scalable (Eklund and Beck 2019); (V) Promote General Data Protection Regulation (GDPR) compliance (Precht et al. 2021) and (VI) ultimately enable embedded supervision in key business processes (Auer 2019).

Because of these unique features, the use of DLT has been studied extensively in central banking, mainly on the topic of Central Bank Digital Currencies (CBDC), Payments clearing and Settlement, Asset transfer and ownership, Audit trail and Regulatory market compliance (Dashkevich et al. 2020), but some authors also focus on supervision, and how to improve automation of regulatory reporting using smart contracts (Auer 2019). IS research in compliance received attention following the financial crisis in 2008 (Abdullah et al. 2009), however, with the influx of new regulations, companies' focus shifted towards meeting regulatory deadlines at the expense of developing a strategic, enterprise-wide connected approach to compliance (Gozman and Currie 2015).While the open-source approach associated with public blockchains was initially opposed by the prevailing thinking in traditional financial services, major institutions on all continents are now experimenting with the technology in view of its attractive characteristics. As a result, banks now represent more than 30 pct of DLT use cases (Rauchs and Hileman 2018), further exacerbated by the evolution of research within machine-readable regulation (McLaughlin and Stover 2021) and business process compliance (López et al. 2020). Yet, in spite of the extensive empirical and theoretical exploration of DLT in the financial settings few concrete results have materialized.

## Methodology and Artefact Requirements

We apply the Design Science Research (DSR) method in an iterative design, develop and evaluation process (Gregor and Hevner 2013). The artefact is conceptualized through a 6-step process: 1) Problem identification, 2) Solution objective, 3) Design, 4) Demonstration, 5) Evaluation and 6) Communication (Ken Peffers et al. 2007). The artefact evaluation process is conducted ex-ante, through expert interviews (Venable et al. 2016), emphasizing the mitigation of development risk through continuous feedback loops (vom Brocke et al. 2020).

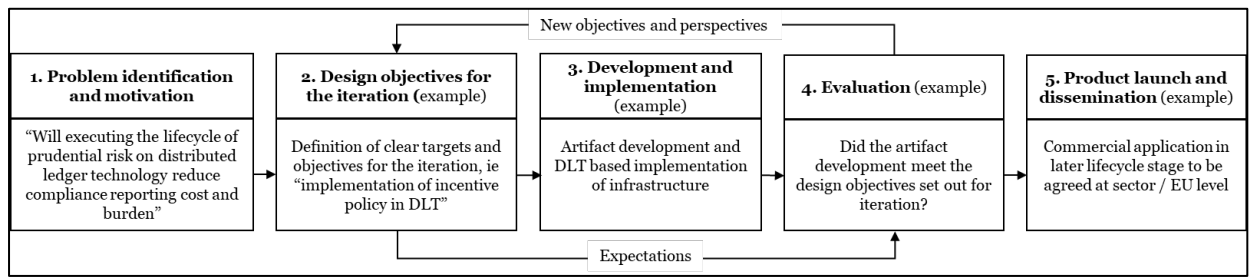

**Figure 3: DSR framework applied to the project's search process**

Our search process was initiated by a 2-day workshop with several of the stakeholders. Here, we identified and motivated the problem and defined the key objectives for the artefact. We subsequently conducted interviews with stakeholders[1] in the period between January to April 2022.

|    | Role in host-organization | Role in the design search process |
|----|---------------------------|-----------------------------------|
| S1 | CEO, Banking Technology Company | Domain expertise |
| S2 | Deputy Director General, Supervisory authority | Domain expertise, guidance, and support |
| S3 | Digital Expert, Central Bank | Non-technical evaluation of artefact requirements |
| S3 | Head of Innovation, Central Bank | Non-technical evaluation of artefact requirements |
| S4 | Solution Architect, Central Bank | Technical evaluation of artefact requirements |
| S5 | Supervisory Data Team, Supervisory authority | Reporting burden expertise, evaluation of artefact |
| S6 | Head of Reporting, Regulator | Reporting burden expertise, evaluation of artefact |
| S7 | Director, Regulator | Domain expertise, guidance and support |
| S8 | Head of Blockchain, Trading platform | Evaluation of artefact, guidance |

**Table 1: Stakeholder categories and role in the design search process**

The interviews were open-ended and semi-structured, lasting 60 minutes on average. Early in the design-search process, we conducted stakeholder interviews without prior briefing. In the evaluation phase, we briefed stakeholders prior to the interviews, keeping them up to date on the latest iteration of the artefact design. The interviews were conducted ensuring proper consent and confidentiality and using a tailored interview guide (Iyamu 2018), covering (1) Introductions; (2) Presentation of the compliance reporting process challenge; (3) Motivation for DLT based compliance reporting; (4) Requirements; (5) Evaluation of artefact and (6) Closing. Thus far, the total interview time with stakeholders comprised 840 minutes, generating 149 pages of interview notes. The project is open-ended, and all stakeholders have agreed to commit time to participate in evaluations for subsequent iterations of the artefact.

While our data sampling strategy aligned with our preconceptions about the feasibility of and challenges within DLT for compliance reporting, we remained open to new theoretical insights and without any preformulated hypothesis (Urquhart et al. 2010). Through the interviews, it became clear how lack of alignment in stakeholder incentives amplified the existing complexity in the identification and motivation of a narrow problem scope, leading to an emphasis on the need for flexibility and modularity in the artefact design. The design-search process is producing an evolving set of requirements for the artefact. In this short paper, we have summarized the technical and governance related set to six requirements in three general categories (Baskerville and Pries-Heje 2010).

|      | Requirement | Description |
|------|-------------|-------------|
| **Data** | Data sources and interoperability | The artefact must demonstrate the reporting flow from reporters to authorities using a *pull-model* system that is interoperable with multiple other non-integrated data sources (synthetic reporting data) |
| | MRER | The system must create machine-executable versions of reporting requirements, that are expressed in a logical and consistent sequence that can be used by a deterministic computing system |

---

[1] To preserve the anonymity of the authors, the organizations in which the stakeholders are employed have been described superficially in this submission.

|  | Security and Privacy | The system must ensure compliance with data privacy regulation such as GDPR, and regulation as relates confidentiality, while also allowing read and write authority as per delegated governance mandate. |
| --- | --- | --- |
| **Decision rights** | Delegation | Risk and obligations must be delegated to system participants, implying the use of one entry point, a simple legally enforceable framework, and smart contract logic with role-based access and identity management, but also institutional involvement. |
| **Accountability** | Data *pull* vs *push* | The system must allow the public authorities to pull the required information directly via the reporting agents for real-time analysis, supervisory review and evaluation and statistical modelling. |
|  | Relevance and Incentivization | The system must feature strong incentivizes for participation, with "opt-in" mechanisms allowing phased entry for participating banks by reducing cost-of-compliance for local banks and institutions. |

**Table 2: Artefact requirements for the presented iteration**

## Artefact description

This early iteration of the artefact design comprises a general database architecture in which transaction events are parsed and enriched with ITS data and subsequently stored for modular ITS report aggregation. The enriched data comprising the fields that make up ITS reports can be *pulled* by regulators from a data warehouse as data is consumed from the DLT environment and enriched, in near real-time. The architecture is rooted in an active node for the targeted DLT environment. The DLT node is running and on-chain event API that listens to native transaction and smart contract events. The on-chain event API is consumed by the 'Composer', a program which observes state changes on the targeted network and records events associated with addresses registered with participating institutions.

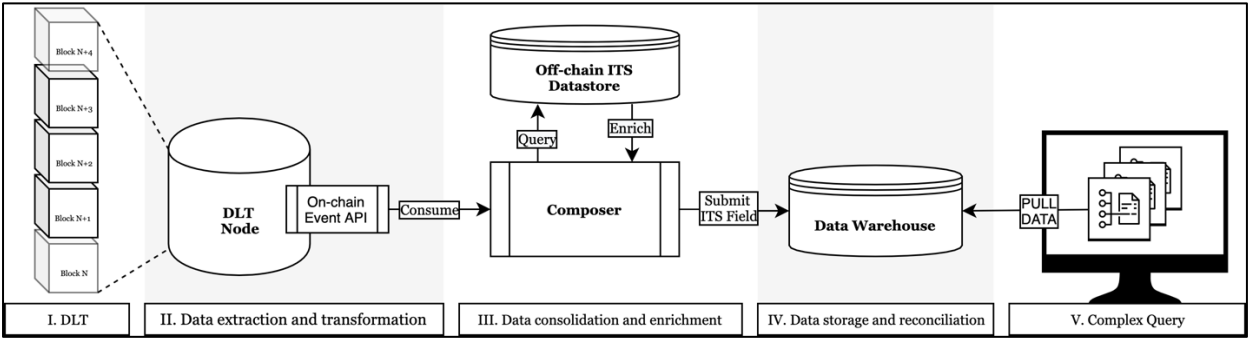

**Figure 4: Artefact illustration – DLT system of systems for compliance reporting**

The Composer queries a DB referred to as the 'ITS Datastore' to enrich the on-chain event data with ITS data and subsequently stores the fields in a data warehouse. The ITS datastore contains relevant information on the institutions operating on the DLT solution, which is used by the Composer in the calculation of leverage and capital ratios, liquidity requirements, credit exposures, trading flows and more. By consuming on-chain events, the Composer maintains logs of activities on the ledger related to participating institutions, which is used in providing a picture of the bank's assets under management.

Regulatory and international bodies can query the data warehouse to pull ITS data as required. The data warehousing is designed such that if enough institutions across jurisdictions would use DLT based solutions, the ITS data can be fed into a nested data hierarchy to compile macro risk assessments and systemic risk analysis in real time, as demonstrated in Table 3.

| **Local Reporting** | **National Reporting** | **Supranational Reporting** |
|---|---|---|
| ITS modular reporting, MI reporting, Idiosyncratic Risk management, Compliance, Strategic direction, Supervision and Evaluation | Predictive local macro, Key Risk Indicators (KRI), Supervision, Systemic country risk, Local financial stability, Prescriptive feedback, Secondary template reporting | Predictive local macro, Key Risk Indicators (KRI), Policy action, Systemic regional risk, Financial stability |

**Table 3: Embedded Data Hierarchy for Real-time ITS Reporting Aggregation**

The ITS supervisory reporting regime requires modular reporting obligations at the primary level – from local banks to national authorities, and further consolidated template reporting obligations in the secondary level – from national authorities to supranational ditto, where the consolidated templates are prepared by the authorities and subject to comprehensive data quality checks to comply with EBA's data point model. As explained earlier, this process is very complicated, expensive and often subject to considerable processing time, which prevents the authorities from dealing precisely with critical matters of financial stability in whole countries in real-time, hence any improvement to the status quo is but important and required, which is also the motivation for EBA's ongoing work and the recent EU supervisory data strategy objectives "to modernize EU supervisory reporting and put in place a system that delivers accurate, consistent, and timely data to supervisory authorities at EU and national level, while minimizing the aggregate reporting burden for all relevant parties" (European Commission 2021b).

## Evaluation

While the choice of working with a broad selection of representatives from industry and regulatory backgrounds has infused the requirements elicitation process with a heterogeneous perspective on compliance and regulation, working in a complex environment with a 'big tent' always introduces discrepancies of opinion and priorities. As a result, the evaluation summary (Table 4) portrays multiple latent discrepancies, as can be expected at this stage in the design-search process.

| | Requirement | Evaluation summary |
|---|---|---|
| **Foundation** | Data sources and interoperability | The artefact conceptually demonstrates the reporting flow at the local reporting level, but allows also secondary reporting from national authorities to supranational ones; it demonstrates how interoperability and enrichment with multiple synthetic reporting data sources will be executed, |
| | MRER | To the extent that regulatory documents and other formal and informal legal documents are enhanced with extensive metadata fields that tell machines and human readers of the types of impacts the document will have, and that the document pertains to, and how restrictive a given document is, the artefact may incorporate machine-readable and executable reporting requirements. |
| **Decision rights** | Security and Privacy | The artefact is designed with the capabilities of DLT as demonstrated by IS scholars above, including compliance with privacy regulation and secure access rights per delegated authority. For security risk, the artefact is intended to comply with prevailing security standards such as ISO 31000/31022 and ISO/IEC 27005, but further work is required to assess the artefact in respect of evolving European cyber security technical standards and resilience testing. |
| | Delegation | A comprehensive regulatory framework does not currently exist for DLT based compliance reporting as outlined, but conditional to the artefact being deployed in a permissioned environment with regulated, identified participants, it will be possible to meet the requirements, assuming tokenized representations are deemed legally enforceable by regulators |
| **Accountability** | Data pull vs push | The artefact demonstrates the feasibility of a "pull" approach, which can hypothetically reduce the time requires for report processing from up to T+90 days towards T+0 days. It is noted that senior management regimes prescribe that management cannot relieve their responsibilities for compliance, regardless of whether a "single source of truth" system is operational. Further work is needed to investigate how this liability regime can be adopted to a DLT system, and how to decide how regulatory breaches are investigated and managed, to avoid triggering undue penalties for wrong calculations or other mistakes in processing the desired outcomes. |
| | Relevance | The artefact governance may allow the incentivization of distributed ownership, delivering single source of truth, that can be utilized at different hierarchical levels. It may also allow increased sharing and standardization of data transformation methods, but further work is required to detail which elements of the ITS is suitable for automation, hence the actual incentive design and regulatory process adaption of DLT technology remains to be completed. |

**Table 4: Artefact requirements for the presented iteration**

The artefact is designed to allow real-time supervisory *pull* access to consolidated and/or granular data, which is very transformational to the current practices. To achieve continued balance in the supervisory review and evaluation process in a system, where supervisors may gain increased level of knowledge of systemic and idiosyncratic risk due the transparent nature of DLT, safeguards may potentially be required to ensure, that bank governance meets other obligations, for example as relates market disclosure. Further, as this is new technology that changes established practices in sensitive areas, it is important to conduct trials in different kinds of settings and with different prototypes prior to production (Lindman et al. 2017).

## Discussion

In this short paper we report ongoing progress on the design of an artefact, with a group of industrial and regulatory stakeholders. The artefact demonstrates the feasibility of implementing a *pull-model* for compliance data, through which regulators can query transaction level reporting data and ultimately stage aggregated financial exposures without disclosing underlying individual transactions. From a supervisory perspective, pulling data directly from banks' ledgers may be perceived not only radical, but counter to tradition because supervision, as it is practiced today, is based on consolidated data, with the intent of understanding the banks' own view of their data. Traditionally, local bank managers interpret data themselves in view of their risk appetite and tolerance, allowing for a lot of flexibility in the calculation of fair value or risk positions. As a result, a *pull-model* may be challenging to operationalize in a secretive industry, where internal control processes are commonly practiced through the '3-lines-of-defence' model, which heavily emphasises the banks' own fiduciary responsibilities (IIA 2020). However, EBA through its consultative process already concluded, that some level of 'pull' is possible, and, as we aim to demonstrate in this short paper, the roll-out and adoption DLT based solutions in the financial services simply advances this principle. To this end, we concur with the existing consensus in the IS literature on DLT and blockchain technologies that is a suitable infrastructure to deliver significant benefits and reduce the compliance reporting burden, while also enabling faster reports processing and decision making. By extrapolating the preliminary findings presented in this paper, we offer the following propositions on the impact of DLT in compliance reporting:

**P1**: DLT based compliance reporting introduces a new level of precision in supervision: The increased level of transparency enables more effective and focused supervision and more precise and faster data sharing across the regulated entities, reducing idiosyncratic and systemic risk. Issues around loss of control, cost of maintaining platform and the risk of intrusive supervision appear more perceived than real (Gozman et al. 2020).

**P2**: Automation through DLT will reduce cost of compliance reporting and improve processing time significantly: The standardization of data taxonomies will lead to increased levels of automation and result in faster and more efficient compliance reporting, reducing cost significantly and eventually paving the way for embedded supervision (Auer 2019).

**P3**: DLT based compliance reporting incentivizes more accurate reporting requirements: As authorities are tasked with creating their own view of banks' data there is a clear incentive for improving the reporting requirements and embrace machine readable regulation (McLaughlin and Stover 2021).

**P4**: DLT will transform how compliance is undertaken: Moving towards a 'pull' model will challenge prevailing control practices such as the '3-lines-of-defence' model, that is widely used for compliance across industries. With increased levels of automation and smarter and more precise reporting requirements, it may not be required to operate what may sometimes in banking behave as a '5-lines-of-defence' model, where the external auditors and authorities are the remaining lines, but with overlaps in between them. Rather, inscription will evolve as the organizing principle, where the existing practices are inscribed in technological artefacts and control is dynamically negotiated (Andersen and Bogusz 2019).

## Conclusions

In this paper we have investigated the complexities of prudential compliance reporting and the implications and potential benefits of using DLT infrastructure for compliance reporting, using the example of the EBA's ITS regime and banking. While DLT applications and the proposed shift to a 'pull' model may present significant change and challenge the prevailing internal control practices, the risks associated with this change are manageable and the benefits clear, as not only will DLT applications pave the way for faster and better reporting at much less cost, DLT will also enable better supervision while incentivizing further automation in a critical, but very expensive, industry. Eventually we predict that as DLT evolves and replaces any firm's legacy technology infrastructure, the reporting burden will not only be possible to reduce significantly, but control activities as such will be optimized via new governance models.